\documentclass[preprint, double-spaced]{revtex4}
\usepackage{wrapfig}
\usepackage{graphicx}
\usepackage{amsmath,bbm}
\usepackage{amssymb,bm}

\begin{document}

\title{Description of Hot and Dense Hadron Gas Properties in a New Excluded-Volume model}
\author{S.~K.~Tiwari\footnote{corresponding author: $sktiwari4bhu@gmail.com$}}
\author{P.~K.~Srivastava}
\author{C.~P.~Singh}

\affiliation{Department of Physics, Banaras Hindu University, 
Varanasi 221005, INDIA}

\begin{abstract}
\noindent
A new equation of state for a hot and dense hadron gas (HG) is obtained where the finite hard-core size of baryons has been incorporated in a thermodynamically consistent formulation of excluded volume correction. Our model differs from other existing approaches on the following points. We assign a hard-core volume only to each baryon and mesons though possess a small volume but they can fuse and interpenetrate into one another. Use of the full quantum statistics is made in obtaining the grand canonical partition function where excluded-volume correction has been incorporated by explicitly integrating over volume. We thus find that the new model works even for the cases of extreme temperatures and/or densities where most of other approaches fail. The model does not violate causality even at extreme densities. The temperature and density dependence of various thermodynamical quantities, e.g. pressure, baryon density, entropy and energy density compare well with the results of other microscopic HG models. After suitable parametrization of the centre-of-mass energy in terms of temperature and baryon chemical potential, we explore some new freeze-out criteria which exhibit full independence of the collision energy and of the structures of the colliding nuclei. We further demonstrate the suitability of our model in explaining various experimental results of the multiplicity-ratios of various particles and their antiparticles. Finally, we use our excluded-volume model to obtain the transport behaviour of the hot and/or dense HG such as shear viscosity to entropy ratio, speed of sound etc. and compare the results with earlier calculations.  
\\


\end{abstract}
\maketitle 
\section{Introduction}
\noindent
One of the most important predictions of quantum chromodynamics (QCD) is regarding a phase transition from hot, dense hadron gas (HG) to a deconfined and/or chiral symmetric plasma of quarks and gluons called as quark-gluon plasma (QGP) which occurs at large temperature and/or baryon density. However, inspite of extensive experimental and/or theoretical research work performed during the last three decades, precise qualitative and quantitative predictions for many aspects of this phase transition are still missing [1-6]. Even the phase diagram for the phase transition is quite uncertain and still exists as a conjectured one. Ultra-relativistic heavy-ion collisions offer the best method to study the properties of QGP in the laboratory. However, the direct observation of the primordial plasma is impossible in the laboratory due to confinement problem. Moreover, QGP survives for a very brief time only and hence after subsequent expansion and cooling, QGP finally hadronizes into a dense and hot HG [7]. Thus QGP diagnostics becomes a very complicated field of study because of our limited knowledge of the HG background. In this context, the search for a proper equation of state (EOS) is of extreme importance because it can suitably describe the properties of hot and dense HG. There are compelling reasons for investigating the properties of HG in unusual environments, in particular at large temperatures and/or baryon densities. The cosmological situations after the big-bang, the interior of the neutron stars and the matter produced in the laboratory after ultra-relativistic heavy-ion collisions are all governed by the presence of such HG matter, and hence a proper EOS can help us to analyze the properties of the matter in above systems.
        In an ideal HG description, all the mesons and baryons are treated as pointlike and non-interacting. However, using Gibb's construction of equilibrium phase transition between HG and QGP, we find an anomalous phase reversal from QGP to HG at large $\mu_B$ and T in the ideal HG picture [8]. This anomalous situation is usually cured by assigning a finite and hard-core volume to each baryon which results in a strong repulsive force between a pair of baryons or anti baryons. Thus any fireball created in a heavy-ion collision at a fixed $T$ and $\mu_B$, cannot accommodate more than a limiting number of baryons because it's volume becomes completely occupied. Moreover, it restricts the mobility of the baryons in the fireball and as a consequence, the thermodynamic pressure of HG is also considerably reduced. Does this kind of 'jamming' also results into the phase transition as we often notice e.g. in the percolation theory ? Therefore, it is worthwhile to study in detail a statistical thermodynamic model in which geometrical hard-core volume of each baryon has been incorporated as excluded-volume effect [8]. The purpose in this paper is to formulate a new thermodynamically consistent excluded-volume model where we assign a finite hard-core volume to each baryon but mesons in the theory can easily overlap, fuse and interpenetrate into each other. Secondly excluded-volume correction has been obtained by performing an explicit integration over 'available' volume in the grand canonical partition function. Thirdly we use full quantum statistics so that our formulation is valid for extreme cases of temperature/density. Our model differs from others mainly on the above features. Here we wish to examine the predictions of our model and make a detailed comparison with the experimental results. We emphasize that we have earlier used this model successfully in obtaining the conjectured QCD phase boundary and thus determining precisely the location of QCD critical end point [9,10]. We have also calculated the freeze-out curve and we notice that the critical end point indeed exists almost on the freeze-out curve. The plan of this paper runs as follows: the ensuing section is devoted to the model description and we have derived our version of the thermodynamically consistent EOS for the hot, dense HG. Then we calculate different thermodynamical quantities like number density, energy density, entropy density, pressure etc. of HG and compare our model calculation with other calculations [11]. In the next section, we analyze the experimental data on the particle multiplicities and ratios for central nucleus-nucleus collision in terms of our model over broad energy range from the lowest GSI Schwerionen Synchrotron (SIS) energy to the highest Relativistic Heavy Ion Collider (RHIC) energies to extract the chemical freeze-out temperatures and baryon chemical potentials which are then suitably parametrized in terms of the centre-of-mass energies and subsequently some chemical freeze-out criteria are also deduced. Thermal fits computed within the statistical models have often been used to successfully reproduce the hadron yield ratios obtained in experimental heavy-ion collisions [12,13,14,15,16,17,18,19,20]. We use our freeze-out picture for calculating the hadron  multiplicities and ratios of strange and non-strange hadrons and compare our results with the experimental data. We also predict the hadron yields which we expect at the LHC energy.  We further use this prescription to calculate pion and nucleon densities and a good comparison between our calculation and Hanbury-Brown-Twiss [HBT] experimental data demonstrate the validity of our model. We also investigate the validity of different freeze-out criteria in our model and conclude that at chemical freeze-out, energy per particle in the fireball $E/N\approx1$ and the entropy per particle $S/N\approx7.0$ emerge as the most appropriate criteria which are almost independent of collision energies and the structures of colliding nuclei. In order to make the discussion complete, we further derive $\eta/s$ and speed of sound from our model and compare with others models. In the concluding section, we focus our attention to the hadron ratios where our model fails and which warrant exotic phenomenon e.g. QGP formation as a suitable alternative to understand them properly.                 

\section{Model Description}
\noindent
We briefly describe our derivation of the EOS for the HG, based on the excluded-volume correction [8,21] where we have assigned a hard-core size to all the baryons while mesons are still treated as pointlike particles in the grand canonical partition function. Moreover, unlike our old paper [8] where Boltzmann statistics make the calculation simple, we use here the full quantum statistics. Thus the grand canonical partition function can be written as follows:

\begin{equation}
\begin{split}
ln Z_i^{ex} = \frac{g_i}{6 \pi^2 T}\int_{V_i^0}^{V-\sum_{j} N_j V_j^0} dV
\\
\int_0^\infty \frac{k^4 dk}{\sqrt{k^2+m_i^2}} \frac1{[exp\left(\frac{E_i - \mu_i}{T}\right)+1]}
\end{split}
\end{equation}
where $g_i$ is the degeneracy factor of ith species of baryons,$E_{i}$ is the energy of the particle ($E_{i}=\sqrt{k^2+m_i^2}$), $V_i^0$ is the eigenvolume assigned to each baryon of ith species and hence $\sum_{j}N_jV_j^0$ becomes the total occupied volume where $N_{j}$ represent the total number of baryons of jth species.

We can clearly write Eq.(1) as:

\begin{equation}
ln Z_i^{ex} = V(1-\sum_jn_j^{ex}V_j^0)I_{i}\lambda_{i},
\end{equation}
where $I_{i}$ represents the integral:
\begin{equation}
I_i=\frac{g_i}{6\pi^2 T}\int_0^\infty \frac{k^4 dk}{\sqrt{k^2+m_i^2}} \frac1{\left[exp(\frac{E_i}{T})+\lambda_i\right]},
\end{equation}
and $\lambda_i = exp(\frac{\mu_i}{T})$ is the fugacity of the particle, $n_i^{ex}$ is the number density after excluded-volume correction and can be obtained from Eq.(2) as :
\begin{equation}
n_i^{ex} = \frac{\lambda_i}{V}\left(\frac{\partial{ln Z_i^{ex}}}{\partial{\lambda_i}}\right)_{T,V}
\end{equation}
Thus our prescription is thermodynamically consistent and it leads to a transcendental equation :
\begin{equation}
n_i^{ex} = (1-R)I_i\lambda_i-I_i\lambda_i^2\frac{\partial{R}}{\partial{\lambda_i}}+\lambda_i^2(1-R)I_i^{'}
\end{equation}
where $I_{i}^{'}$ is the partial derivative of $I_{i}$ with respect to $\lambda_{i}$and $R=\sum_in_i^{ex}V_i^0$ is the fractional occupied volume. We can write R in an operator equation form as follows [9]:
\begin{equation}
R=R_{1}+\hat{\Omega} R
\end{equation}
where $R_{1}=\frac{R^0}{1+R^0}$ with $R^0 = \sum n_i^0V_i^0 + \sum I_i^{'}V_i^0\lambda_i^2$; $n_i^0$ is the density of pointlike baryons of ith species and the operator $\hat{\Omega}$ has the form :
\begin{equation}
\hat{\Omega} = -\frac{1}{1+R^0}\sum_i n_i^0V_i^0\lambda_i\frac{\partial}{\partial{\lambda_i}}
\end{equation}
Using Neumann iteration method and retaining the series upto $\hat{\Omega}^2$ term, we get
\begin{equation}
R=R_{1}+\hat{\Omega}R_{1} +\hat{\Omega}^{2}R_{1}
\end{equation}
\noindent
After solving Eq.(8), we finally get the expression of total pressure [21] for the hadron gas as:
\begin{equation}
P^{ex} = T(1-R)\sum_iI_i\lambda_i + \sum_iP_i^{meson}.
\end{equation}
Here $P_i^{meson}$ is the  pressure due to ith type of mesons.

In Eq.(9), the first term represents the pressure due to all types of baryons where excluded-volume correction is incorporated and the second term gives the pressure arising due to all mesons in HG as if they possess a pointlike size. This makes it clear that we consider the repulsion arising only between a pair of baryons and/or antibaryons because we assign them exclusively a hard-core volume. In order to make the calculation simple, we have taken an equal volume $V^{0}=4\pi r^{3}/3$ for each type of baryons with a hard-core radius $r=0.8\;fm$. We have considered in our calculation all baryons and mesons and their resonances having masses upto a cut-off value of $2\;GeV/c^{2}$ and lying in the HG spectrum. Here resonances having well defined masses and widths have only been incorporated in the calculations. Branching ratios for sequential decays have been suitably accounted and in the presence of several decay channels, only dominant mode is included. We have also imposed strictly the condition of strangeness neutrality by putting $\sum_{i}S_{i}(n_{i}^{s}-\bar{n}_{i}^{s})=0$, where $S_{i}$ is the strangeness quantum number of the ith hadron, and $n_{i}^{s}(\bar{n}_{i}^{s})$ is the strange (anti-strange) hadron density, respectively. Using this constraint equation, we get the value of strange chemical potential in terms of $\mu_{B}$. Having done all these things, we proceed to calculate the energy density of each baryon species i by using the following formula :
\begin{equation}
\epsilon_{i}^{ex}=\frac{T^{2}}{V}\frac{\partial lnZ_i^{ex}}{\partial T}+\mu_{i} n_{i}^{ex} 
\end{equation}
Similarly entropy density is:
\begin{equation}
s=\frac{\epsilon_{i}^{ex}+P^{ex}-\mu_{B}n_{B}-\mu_{S} n_{S}}{T}
\end{equation}

It is evident that this approach is more simple in comparison to other thermodynamically consistent, excluded-volume approaches which often possess transcendental final expressions and are found them usually difficult to solve [14]. This approach does not involve any arbitrary parameter in the calculation. Moreover, this approach can be used for extremely low as well as extremely large values of $T$ and $\mu_B$  where all other approaches fail to give a satisfying result [14].

\section{Hadronic Properties}

\begin{figure}
\includegraphics[height=20em]{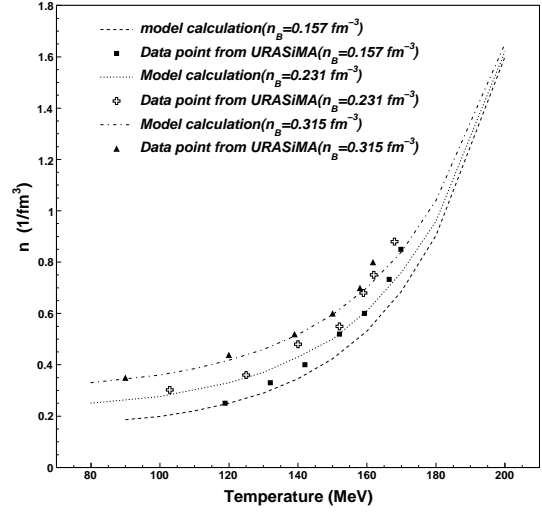}%use pdflatex for pdf
\caption[]{Variation of total number density with respect to temperature at constant net baryon density. Solid lines show our model calculation and solid points are the data calculated by Sasaki using URASiMA event generator.}
\end{figure}

\begin{figure}
\includegraphics[height=20em]{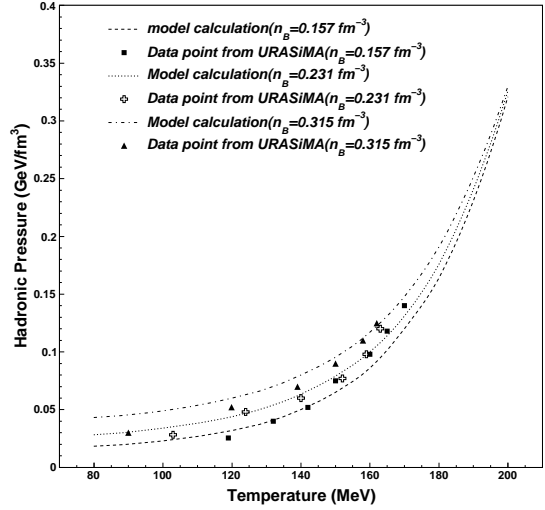}%use pdflatex for pdf
\caption[]{Variation of pressure with respect to temperature at constant net baryon density. Solid lines show our model calculation and solid points are the data calculated by Sasaki using URASiMA event generator}
\end{figure}

In this section, we attempt to calculate the number density, pressure, energy density and entropy density of hadrons and compare the results with the predictions of a microscopic model named as URASiMA generator [11] which is essentially based on the molecular-dynamical simulation performed for a system of hadrons. In Fig. 1, we have shown the variation of total number density of hadrons with respect to the temperature at fixed baryon density and compared with the results obtained by URASiMA event-generator. The results show a very close agreement between our model calculation and the results of Sasaki but at higher T, the curves seem to differ slightly.  

In Fig. 2, we have plotted the variation of total pressure generated by all the hadrons with respect to temperature at fixed net baryon density. Hadronic pressure initially shows a very slow increase but for $T\geq\;170 MeV$, the pressure increases rapidly. The hadronic pressure calculated in our model again shows a good agreement with the results of Sasaki [11]. It shows that the EOS of HG as given by our excluded-volume model incorporating macroscopic geometrical features gives results in close agreement with the simulation involving hadrons and hadronic interactions. The method of Sasaki [11] involves various parameters, e.g., coupling constants of hadrons etc. arising due to hadronic interactions. However, it is encouraging to see such excellent matching of the results obtained with two  widely different models. 

\begin{figure}
\includegraphics[height=20em]{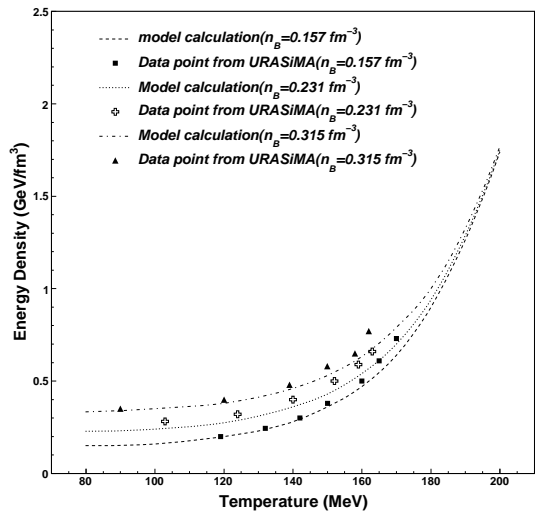}%use pdflatex for pdf
\caption[]{Variation of energy density with respect to temperature at constant net baryon density. Solid lines show our model calculation and solid points are the data calculated by Sasaki using URASiMA event generator}
\end{figure}

\begin{figure}
\includegraphics[height=20em]{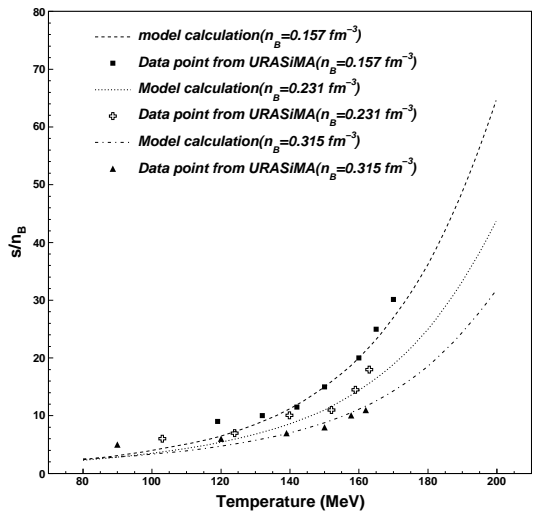}%use pdflatex for pdf
\caption[]{Variation of $s/n_B$ with respect to temperature at constant net baryon density. Solid lines show our model calculation and solid points are the data calculated by Sasaki using URASiMA event generator}
\end{figure}

Fig. 3 represents the variation of the energy density of HG with respect to temperature at constant net baryon density. Again a very good agreement between our model calculations and the results from URASiMA event generator demonstrates the validity of our model in describing the properties of hot, dense HG. Energy density increases very slowly with the temperature initially and then rapidly increases at higher temperatures. Similarly in Fig. 4, we have shown the variation of entropy per baryon $s/n_B$ in the HG with respect to the temperature at fixed net baryon density. We stress that $s/n_B$ measures the yield of all particles relative to nucleons [22] and ideal HG model gives $s/n_B$ as almost constant quantity for any fireball which means that it does not change during the expansion or evolution of the fireball. So this is a measurable quantity and is significant in fixing the properties of the HG. Again Fig. 4 demonstrates a very good agreement between two models. It should be stressed here that the essential difference between our present model and the earlier calculation [8] is the use of full quantum statistics here in comparison to Boltzmann statistics used in Ref. [8]. We notice that this modification has improved the fit between our model and Sasaki's calculation. Both the models predict a rapid increase in $s/n_B$ for $T\geq\;160 MeV$ even at a fixed $n_B$.

In order to relate the thermal parameters of hot, dense HG with the centre-of-mass energy, we extract them by fitting the experimental particle-ratios from the lowest SIS energy to the highest RHIC energy by our model calculation. We thus deduce the temperature and baryon chemical potential thermodynamically from the experiments at various energies as tabulated in table I. For comparison, we have also shown the values obtained in other models, e.g., Ideal hadron gas (IHG) and Rischke, Gorenstein, St$\ddot{o}$cker, Greiner (RGSG) model. We then parametrize the variables $T$ and $\mu_B$ in terms of centre-of-mass energy as follows [23]: 
\begin{equation}
\mu_B=\frac{a}{1+b\sqrt{s_{NN}}}
\end{equation}
\begin{equation}
T=c-d\mu_B^2-e\mu_B^4 
\end{equation}
where the parameters $a$,$b$,$c$,$d$ and $e$ have been determined from the best fit : $a=1.482\pm0.0037$ $GeV$,$ b=0.3517\pm0.009$ ${GeV}^{-1}$, $c=0.163\pm0.0021$ $GeV$,$ d=0.170\pm0.02$ ${GeV}^{-1}$ and $ e=0.015\pm0.01$ ${GeV}^{-3}$.

\begin{wraptable}{c}{15 cm} 
%\begin{wraptable}
\caption{Thermal parameters $(T,\;\mu_B)$ values obtained by fitting the experimental particle-ratios in different model calculations.}
\begin{tabular}{l|l|l|l|l|l|l|l|l|l|l|l|l}
\hline
 \boldmath{$\sqrt{S_{NN}}$}\textbf{(GeV)} & \multicolumn{3}{l|}{\textbf{IHG Model}}&\multicolumn{3}{l|}{\textbf{RGSG Model}} &\multicolumn{3}{l|}{\textbf{Our Old Model}}  &\multicolumn{3}{l}{\textbf{Our Present Model}} \\
\cline{2-13}
 &\textbf{T}&\boldsymbol{$\mu_{B}$}&\boldsymbol{$\delta^{2}$}&\textbf{T}&\boldsymbol{$\mu_{B}$}&\boldsymbol{$\delta^{2}$}&\textbf{T}&\boldsymbol{$\mu_{B}$}&\boldsymbol{$\delta^{2}$}&\textbf{T}&\boldsymbol{$\mu_{B}$}&\boldsymbol{$\delta^{2}$}\\
\hline\hline

2.70  & 60     &740      & 0.85       & 60      & 740    & 0.75           & 60     & 740        & 0.87       & 70     & 760        & 1.15\\
3.32  & 80     &670      & 0.89       & 78      & 680    & 0.34           & 90     & 670        & 0.69       & 90     & 670        & 0.45\\
3.84  & 100    &645      & 0.50       & 86      & 640    & 0.90           & 100    & 650        & 0.60       & 100    & 640        & 0.34\\
4.32  & 101    &590      & 0.70       & 100     & 590    & 0.98           & 101    & 600        & 0.53       & 105    & 600        & 0.23\\
8.76  & 140    &380      & 0.45       & 145     & 406    & 0.62           & 140    & 380        & 0.26       & 140    & 360        & 0.25\\
12.3  & 148    &300      & 0.31       & 150     & 298    & 0.71           & 148.6  & 300        & 0.31       & 150    & 276        & 0.20\\
17.3  & 160    &255      & 0.25       & 160     & 240    & 0.62           & 160.6  & 250.6      & 0.21       & 155    & 206        & 0.27\\
130   & 172.3  &35.53    & 0.10       & 165.5   & 38     & 0.54           & 172.3  & 28         & 0.056      & 163.5  & 32         & 0.05\\
200   & 172.3  & 23.53   & 0.065      & 165.5   & 25     & 0.60           & 172.3  & 20         & 0.043      & 164    & 20         & 0.05\\ \hline

\hline

\end{tabular}

\end{wraptable}

In this exercise we have taken the experimental data measured in full phase-space (4$\pi$) so that we can remove any possible influence on particle ratios arising due to hydrodynamical flow [24]. This allows us to study the hadronic ratios without bothering about the expansion of the system at freeze-out. 

\begin{figure}
\includegraphics[height=18em]{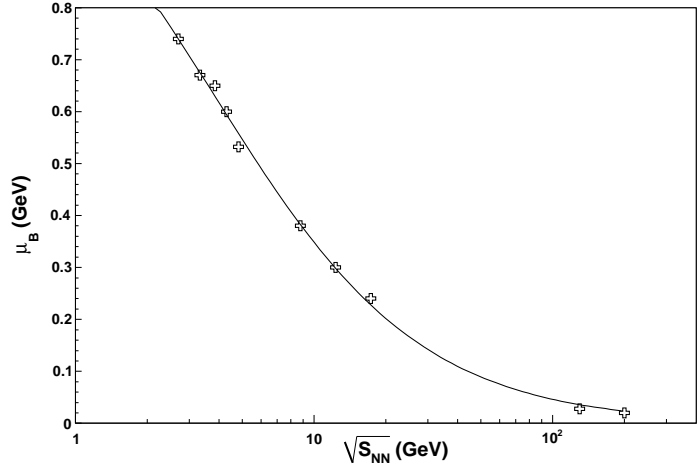}%use pdflatex for pdf
\caption[]{Variation of chemical potential with respect to centre of mass energy}
\end{figure}

\begin{figure}
\includegraphics[height=18em]{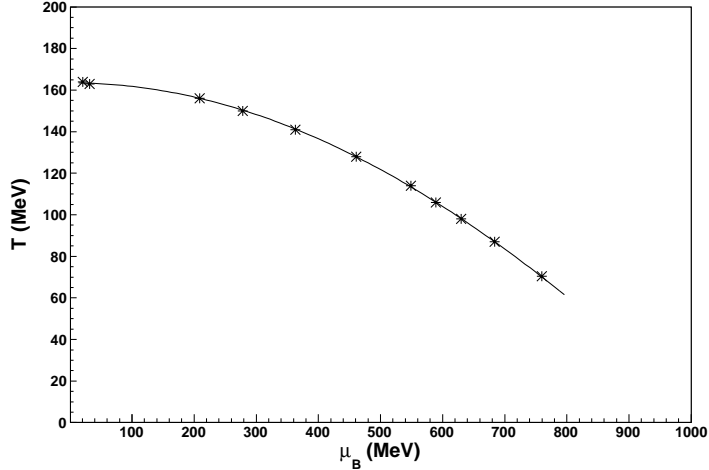}%use pdflatex for pdf
\caption[]{Variation of chemical freeze-out temperature with respect to baryon chemical potential}
\end{figure}

In Fig. 5, we have shown the parametrization of the freeze-out values of baryon chemical potential with respect to the centre-of-mass energy and similarly in Fig. 6, we have shown the chemical freeze-out curve between temperature and baryon chemical potential. The fits demonstrate that the parameters in the parameterizations $(12)$ and $(13)$ have been suitably chosen and the experimental variable such as centre-of-mass energy can be described well by the variables T and $\mu_B$ of the fireball.

\section{The energy dependence of hadron ratios}

\begin{figure}
\includegraphics[height=20em]{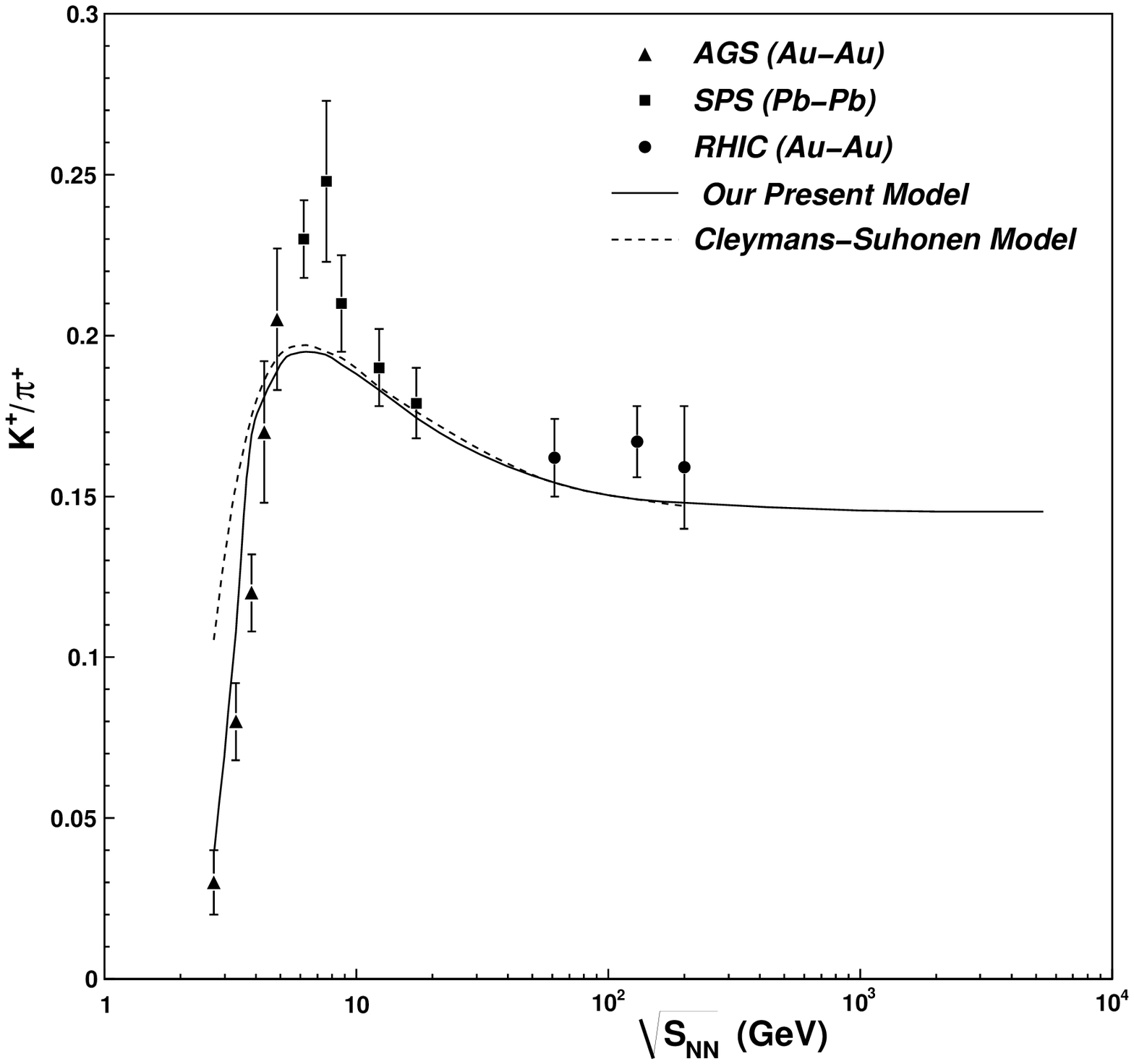}%use pdflatex for pdf
\caption[]{The energy dependence of kaon relative to pion. We have compared our results with the Cleymans-Suhonen model [25]. Solid points are the experimental data [26-28].}
\end{figure} 

\begin{figure}
\includegraphics[height=20em]{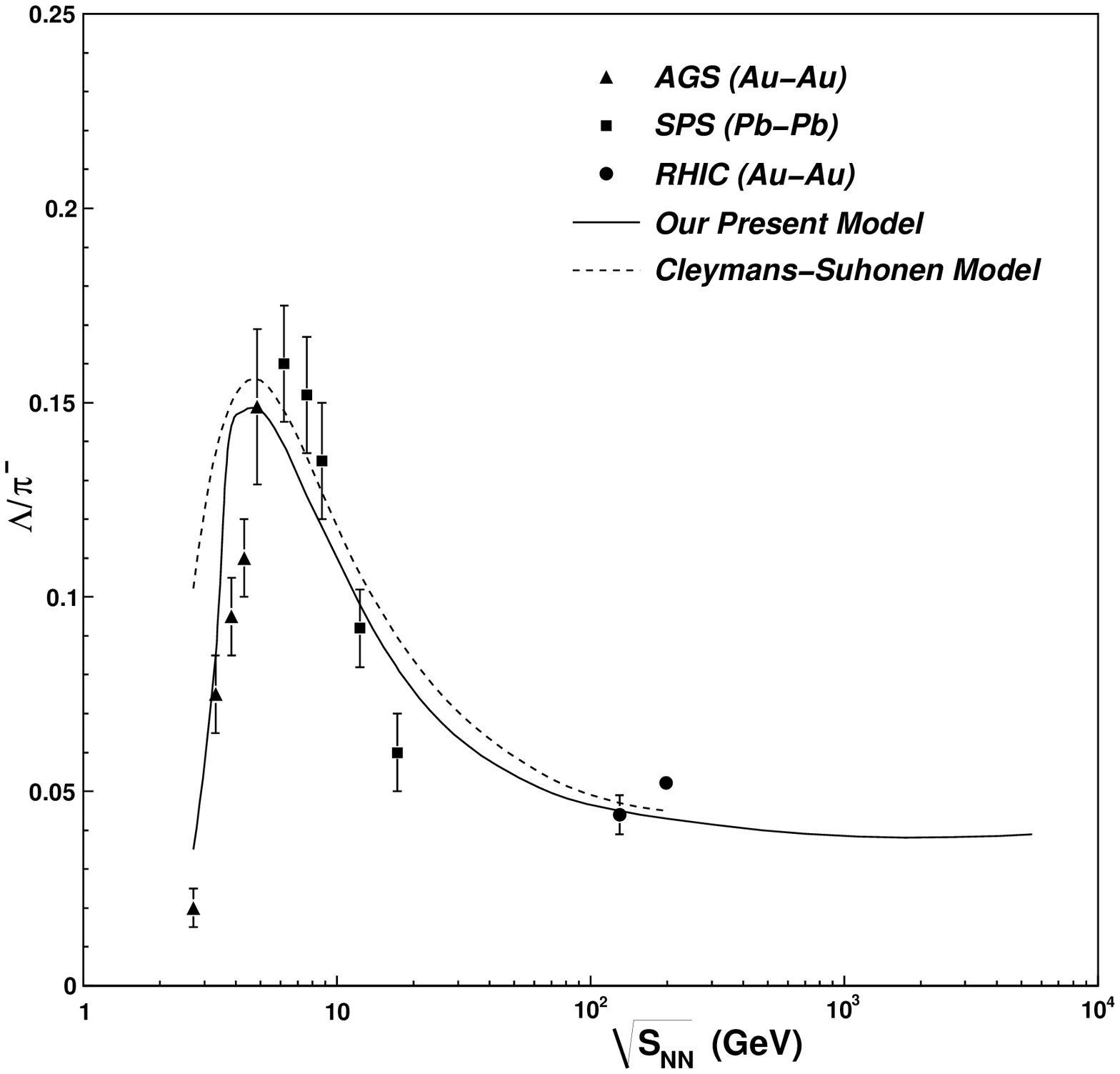}%use pdflatex for pdf
\caption[]{The energy dependence of lambda relative to pion. We have compared our results with Cleymans-Suhonen model [25]. Solid points are the experimental data [26-28].}
\end{figure}

\begin{figure}
\includegraphics[height=20em]{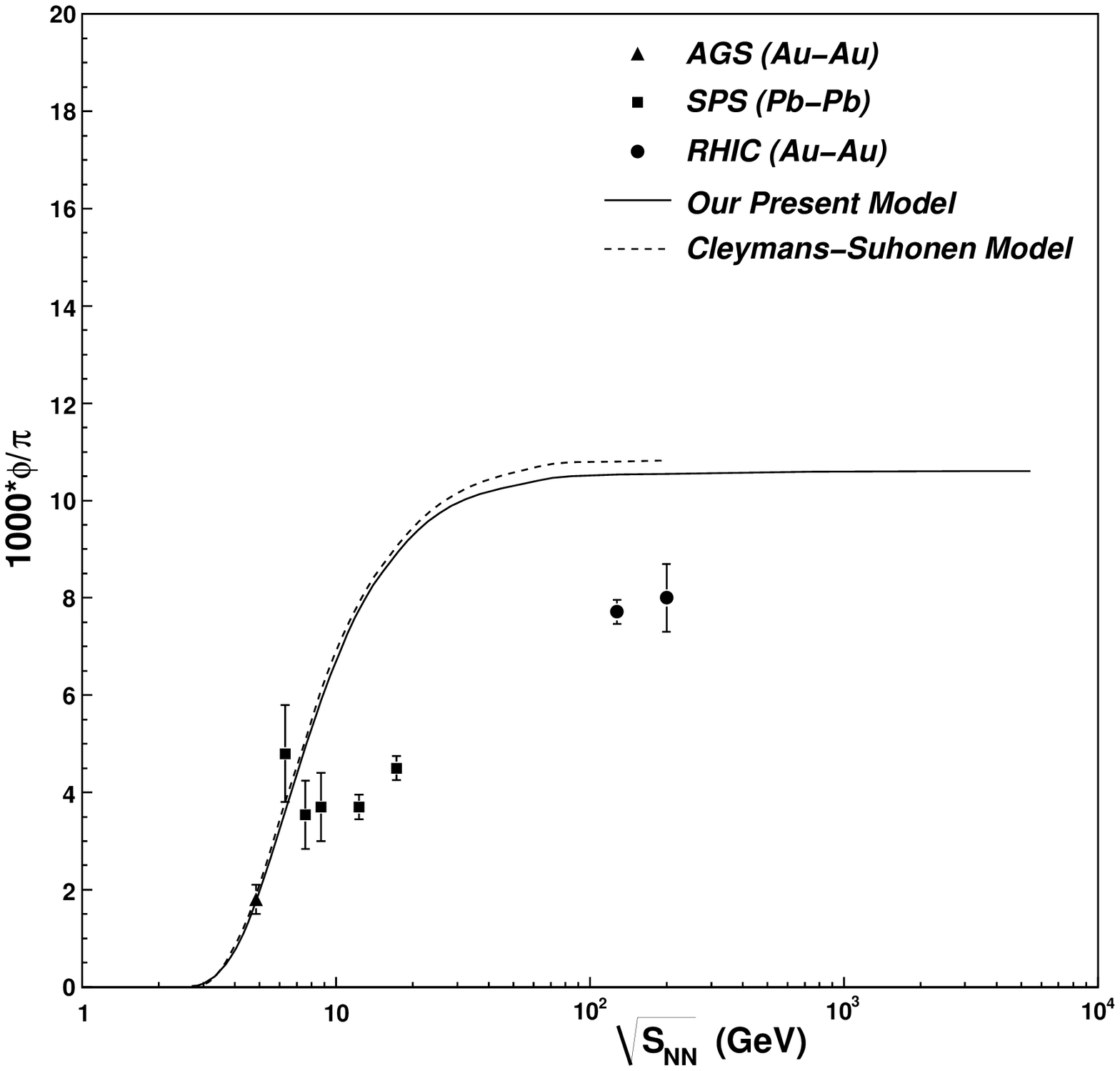}%use pdflatex for pdf
\caption[]{The energy dependence of phi relative to pion. We have compared our results with the Cleymans-Suhonen model [25]. Solid points are the experimental data [26-28].}
\end{figure}

\begin{figure}
\includegraphics[height=20em]{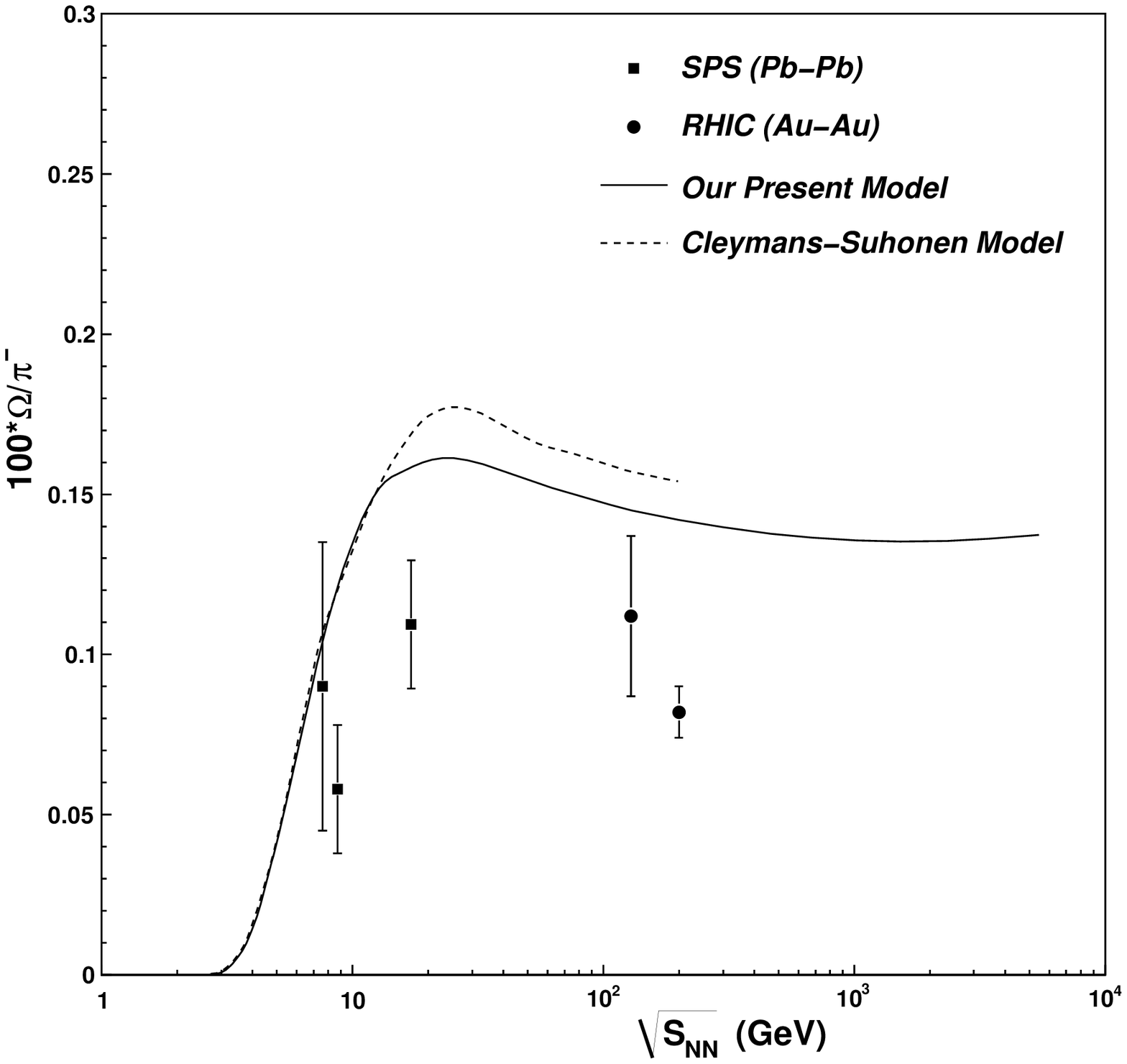}%use pdflatex for pdf
\caption[]{The energy dependence of omega relative to pion. We have compared our results with the Cleymans-Suhonen model [25]. Solid points are the experimental data [26-28].}
\end{figure}

In a series of measurements of Pb-Pb collisions at various centre-of-mass energies [29,30,31,32,33], it is found that there is an unusual sharp variation giving rise to peaks in the $K^{+}/\pi^{+}$ and $\Lambda/\pi^{-}$ ratios. Such a strong variation of $K^{+}/\pi^{+}$  with energy does not occur in p-p collisions and, therefore, has been attributed to the presence of unusual phenomena of the QGP formation. This transition has been referred as ``horn'' in Ref. [29]. A strong variation of $\Lambda/\pi^{-}$ with energy has also been attributed as a signal for the existence of a critical point in the QCD phase diagram [34,35] and a nontrivial information about the critical temperature $T_C\approx176\;MeV$ has been extracted [35]. A sharp rise at low energies with a mild maximum and a subsequent flattening of $K^{+}/\pi^{+}$ was also reported by many authors [15,18,36] using various statistical model calculations. Nayak et al. [37] have also explained the ``horn'' by using a microscopic approach for the HG. Similarly a good fit with the experimental data for the horn has been proclaimed as the onset of  QGP formation [38,39,40]. In Fig. 7, we have shown the results of our calculation for $K^{+}/\pi^{+}$ and we have compared our results with those from other model. We find that our results almost coincide with the results of Cleymans-Suhonen model which involves a thermodynamical inconsistency. Fig. 8 shows the variation of $\Lambda/\pi^{-}$ with $\sqrt{S_{NN}}$. We have again compared our results with various HG models [25] and we find that our model calculation gives much better fit to the experimental data at all energies in comparison to other models. Although we have not successfully reproduced the sharp peak in  $K^{+}/\pi^{+}$ but still we get a broad peak and our results almost reproduce the data at lower as well as higher energies. In $\Lambda/\pi^{-}$ case we get a sharp peak around centre-of-mass energy of 5 GeV and our results almost reproduce all the features of the experimental data.

\begin{figure}
\includegraphics[height=25em]{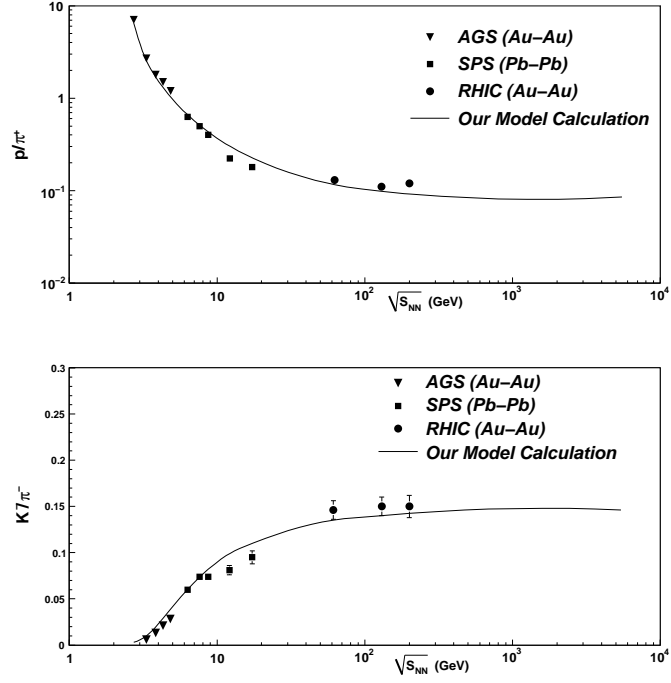}%use pdflatex for pdf
\caption[]{The energy dependence of various hadrons relative to pion. Solid points are from the experimental data [26-28] and solid line represents our model calculation. }
\end{figure}

\begin{figure}
\includegraphics[height=25em]{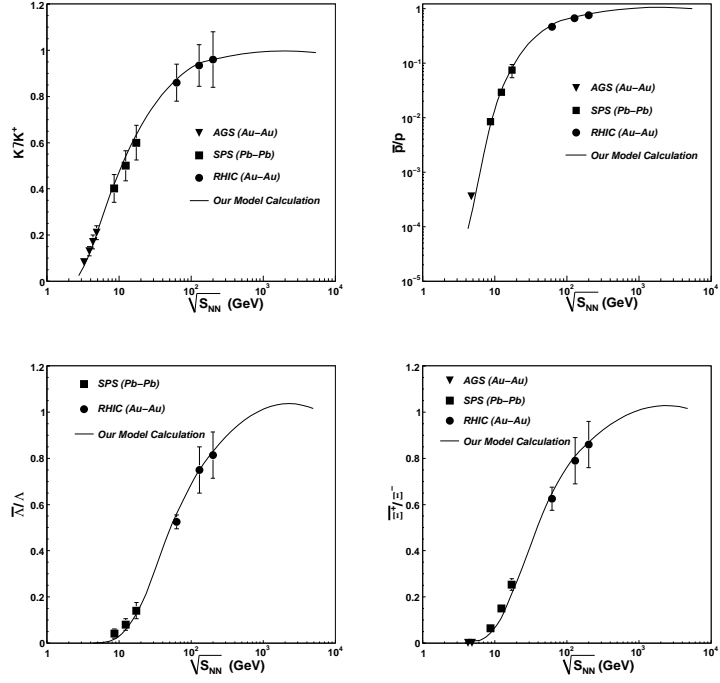}%use pdflatex for pdf
\caption[]{The energy dependence of anti hadron to hadron ratios. Solid points are from the experimental data [26-28] and solid line represents our model calculation.}
\end{figure}

\begin{figure}
\includegraphics[height=20em]{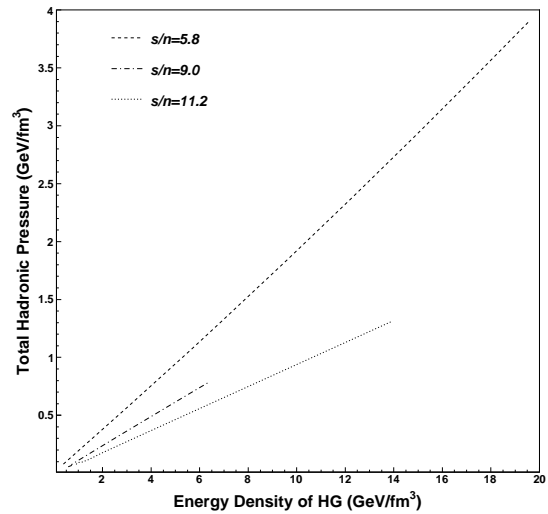}%use pdflatex for pdf
\caption[]{Variations of total hadronic pressure with respect to energy density of HG at fixed entropy per particle $s/n$. Our calculations show linear relationship and slope of the lines give square of the velocity of sound ${c_{s}}^{2}$.}
\end{figure}

\begin{figure}
\includegraphics[height=20em]{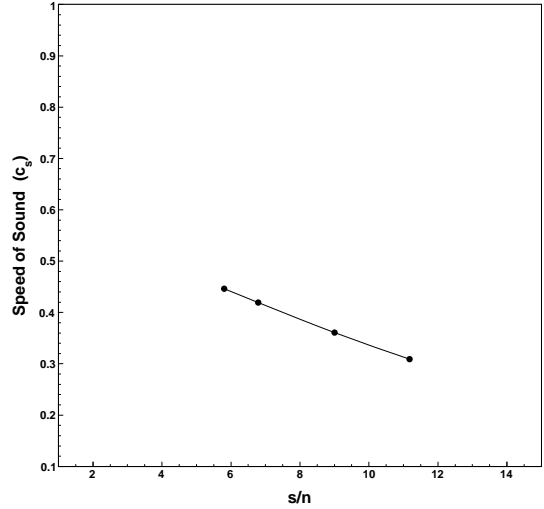}%use pdflatex for pdf
\caption[]{Variation of velocity of sound in the hot, dense HG medium with respect to entropy per particle $s/n$.}
\end{figure}

\begin{figure}
\includegraphics[height=15em]{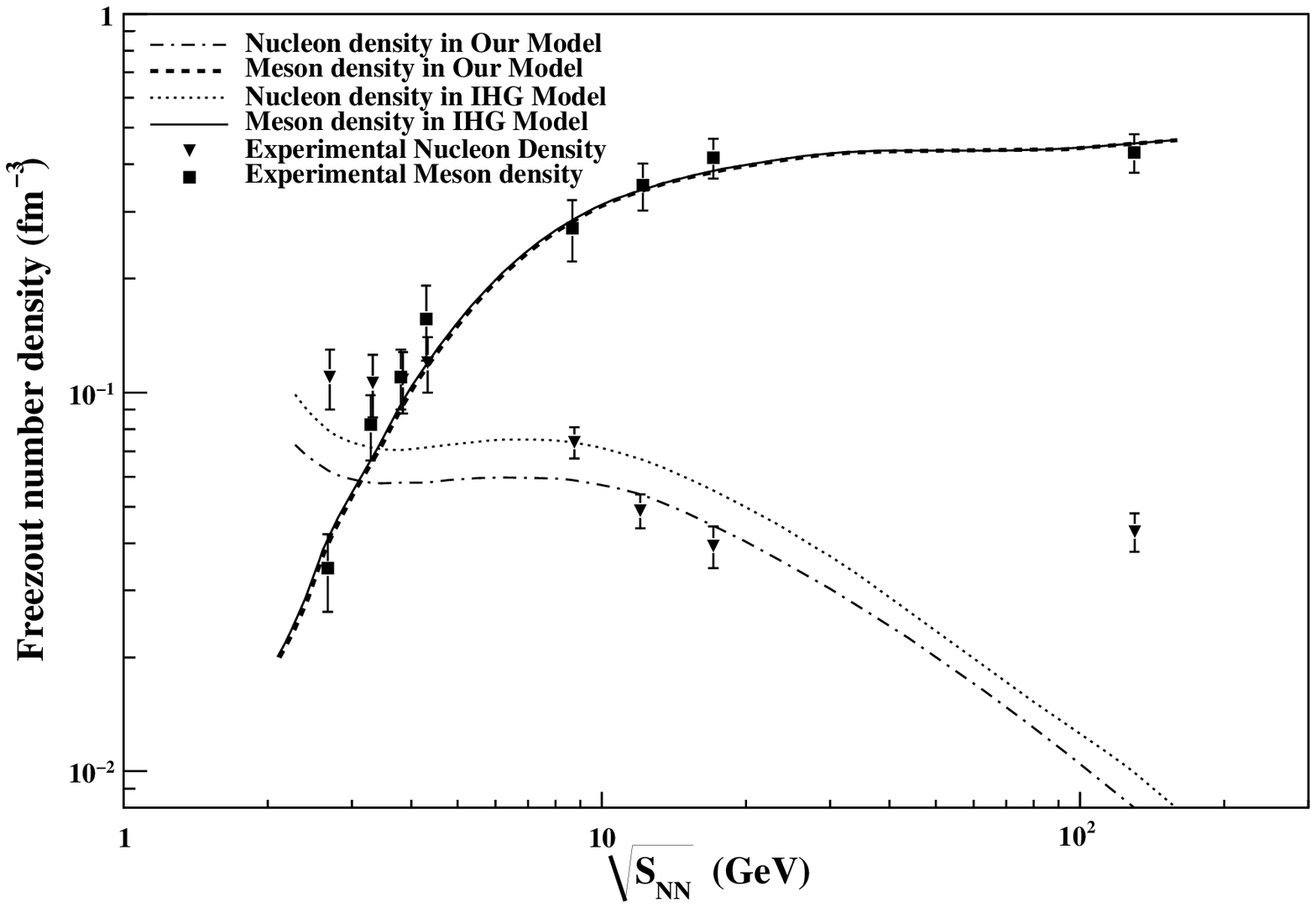}%use pdflatex for pdf
\caption[]{Variation of nucleon density and pion density with respect to centre-of-mass energy $(\sqrt{S_{NN}})$. Solid line shows the meson density calculated in IHG model and dashed line shows our model calculation for meson density. The dotted and dash-dotted lines show the nucleon density calculated in IHG model and our model, respectively. Solid points show the HBT experimental data [42].}
\end{figure}

    In Fig. 9 and Fig. 10, we have shown the variations of the multiplicity-ratios of $\phi$ and $\Omega^-$ relative to pions with the centre-of-mass energy, respectively. Our model is able to reproduce the experimental data only at lower $\sqrt{S_{NN}}$. Although our model calculation is not able to describe these ratios, but it is more closer to the experimental data in comparison to other model specially at higher $\sqrt{S_{NN}}$. We notice that no thermal model can suitably account for the multiplicity-ratios of multi strange particles since $\Omega^-$ is $sss$ and $\phi$ is $s\bar{s}$ hidden-strange quark combinations. Strangeness enhancement invoked in the case of QGP formation will also give the unmatching results. However, quark coalescence model assuming a QGP formation has been claimed to explain the results [41]. In thermal model, this result for the multistrange particles raises doubt over the degree of chemical equilibration for strange hadrons reached in the HG fireball. The failures of excluded-volume models in these cases may indicate the presence of QGP formation. Fig. 11 shows the energy dependence of $K^{-}$ and p relative to pions. There is a very good agreement between our model calculations and the experimental data. These ratios saturate at higher energies which means that the production rate of these particles is independent of $\sqrt{S_{NN}}$ at higher energies. In Fig. 12, we have shown the energy dependence of antiparticle to particle ratios e.g. $K^-/K^+$, $\bar{p}/p$, $\bar{\Lambda}/\Lambda$, and $\bar{\Xi^{+}}/\Xi{-}$. These ratios increase sharply with respect to  $\sqrt{S_{NN}}$ and then almost saturate at higher energies reaching the value equal to 1.0 at LHC energy. This behaviour shows that the production rates of anti-particle relative to particle continuously increase with increasing $\sqrt{S_{NN}}$ and will become almost equal at LHC energy. The excellent agreements between our model results and the experimental data demonstrate the validity of our model in describing the data starting from the lowest upto the highest energy.

Usually the excluded-volume models suffer from a severe deficiency caused by the violation of causality in the hot and dense hadron gas i.e., the sound velocity $c_s$ is larger than the velocity of light $c$ in the medium. In other words, $c_s>1$ in the unit of $c=1$, means that the medium transmits information at a speed faster than $c$ [42]. It would be interesting to see if our model violates causality. In Fig. 13, we have plotted the variations of the total hadronic pressure $P$ as a function of the energy density $\epsilon$ of the HG at a fixed entropy per particle. We find for a fixed $s/n$, the pressure varies linearly with respect to energy density. In Fig. 14, we have shown the variation of $c_s$ (${c_s}^2=\partial P/\partial\epsilon$ at fixed $s/n$) with respect to $s/n$. We find that always $c_s\leq0.58$ in our model of interacting particles having a hard-core size. We get $c_s=0.58$ (i.e. $1/\sqrt{3}$) for an ideal gas consisting of ultra-relativistic particles. This feature endorses our viewpoint that our model is not only thermodynamically consistent but it does not involve any violation of causality.

In Fig. 15, we have shown the variations of nucleon density and pion density with respect to centre-of-mass energy. We have compared our results with IHG model and find that both results are in better agreement with the experimental data [42] as obtained by HBT interferometry [43]. In fact for pion density, we find that the incorporation of hard-core size does not produce any noticeable change. However, for the nucleon-density, we notice that our calculations yield results lying well below the HBT results at lower centre-of-mass energies. Similarly the experimental value for the nucleon density at RHIC energy lies well above our theoretical result. In general, the experimental data for enhanced nucleon-density obtained at recent heavy-ion colliders experiments, have posed a problem which defies explanation. Hence, some other production mechanism is needed to explain the excess of baryon density observed in these experiments.

\section{freeze-out criteria- Revisited}

\begin{figure}
\includegraphics[height=20em]{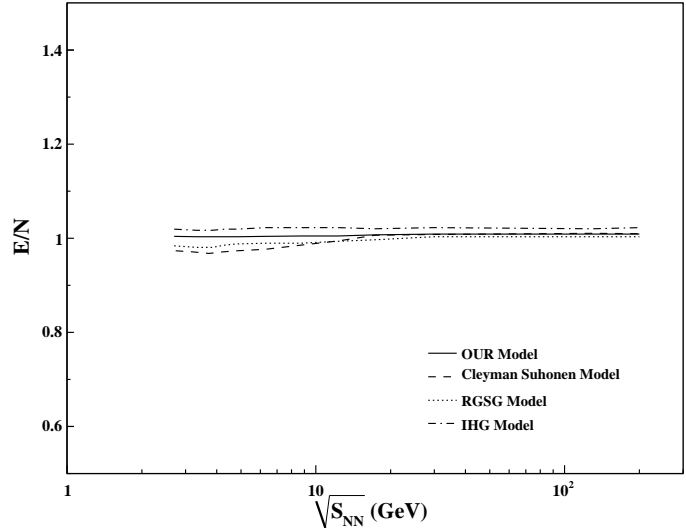}%use pdflatex for pdf
\caption[]{Variation of $E/N$ with $\sqrt{S_{NN}}$. Ideal HG model calculation is shown by dash dotted line,Cleymans-Suhonen and Rischke et al. model (RGSG model) calculation shown by dashed and dotted line respectively. Solid line shows the calculation by our model}
\end{figure}

In ultra-relativistic nucleus-nucleus collisions, a hot, dense matter is formed over an extended region for a very brief time and it is often called a 'fireball'. The physical variables of the fireball are volume $V$, energy density $\epsilon$, and baryon density $n_B$ which are in fact related to $T$ and $\mu_B$ of the fireball. When cooling or expansion of the fireball starts, it goes through two types of freeze-out stages, when inelastic collisions between constituents of the fireball do not occur, we call this as chemical freeze-out stage. Later when elastic collisions also cease to happen in the fireball, this stage specifies the thermal freeze-out. Abundances of particles and their ratios provide important information regarding the chemical equilibrium occurring in the fireball just before the thermal equilibrium.

After seeing the remarkable success of our model in explaining the multiplicities and the particle ratios of various particles produced in heavy-ion experiments from the lowest SIS energy upto the highest RHIC energies, we wish to extend the search of chemical freeze-out criteria for the fireball. Recently many papers have appeared [14,23,44,45,46,47] which predict following empirical conditions to be valid on the entire freeze-out hypersurface of the fireball : $(i)$ energy per hadron is a fixed value at $1.08\;GeV$, $(ii)$ sum of baryon and anti-baryon densities $n_{B}+n_{\bar{B}}=0.12/{fm}^3$, $(iii)$ normalized entropy density $s/T^{3}\approx7$. Indeed Cleymans et al. have found that these conditions separately give a satisfactory description of the chemical freeze-out coordinates of $T$ and $\mu_B$ in an IHG picture of statistical thermodynamics. Moreover, it was also proposed that these conditions are independent of collision energy and the geometry of colliding nuclei but these findings were not illustrated explicitly. Furthermore, Cleymans et al. [23] have hinted that incorporation of excluded-volume correction leads to wild as well as disastrous effects on these conditions. The purpose in this section is to reinvestigate the validity of these freeze-out criteria in our excluded-volume model. 

\begin{figure}
\includegraphics[height=18em]{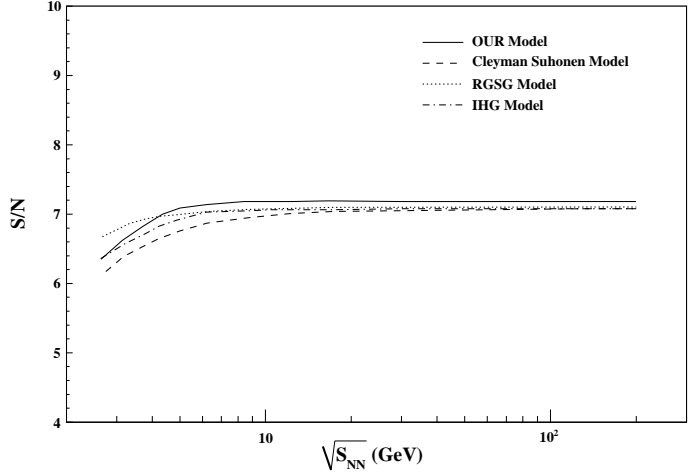}%use pdflatex for pdf
\caption[]{Variation of $S/N$ with $\sqrt{S_{NN}}$. IHG model calculation is shown by dash dotted line,Cleymans-Suhonen and Rischke et al. model (RGSG model) calculation shown by dashed and dotted line respectively.Solid line shows the calculation by our model}
\end{figure}

In Fig. 16, we have shown the variation of $E/N$ with respect to centre of mass energy $(\sqrt{S_{NN}})$ at the chemical freeze-out point of the fireball. The ratio $E/N$ shows indeed a constant value of 1.0 in our calculation and it shows a remarkable energy independence. Similarly the curve in IHG model shows that the value for $E/N$ is slightly larger than one as reported in [23]. However, our results support that $E/N$ is almost independent of energy and also of the geometry of the nuclei. Most importantly we notice that the inclusion of the excluded-volume correction does not change the result much which is contrary to the claim of Cleymans et. al. [23]. The condition $E/N\approx1.0 GeV$ was successfully used in the literature to make predictions [48] of freeze-out parameters at SPS energies of 40 and 80 A GeV for Pb-Pb collisions long before the data were taken. Moreover, we have also shown in Fig. 16, the curves in the Cleymans-Suhonen model [25] and the RGSG excluded volume model [41] and we notice a small variation with $\sqrt{S_{NN}}$ particularly at lower energies.

In Fig. 17, we study a possible new freeze-out criterion which was not proposed earlier. We show that the quantity entropy per particle i.e. $S/N$ yields a remarkable energy independence in our model calculation. The quantity $S/N\approx7.0$ describes the chemical freeze-out criteria and is almost independent of the centre-of-mass energy in our model calculation. However, the results below $\sqrt{S_{NN}}=6\;GeV$ do not give promising support to our criterion and show some energy-dependence. This criterion thus indicates that the possible use of excluded-volume models and the thermal descriptions at very low energies is not valid for the HG. Similar results were obtained in the RGSG, Cleymans-Suhonen model and IHG model also.  

\begin{figure}
\includegraphics[height=20em]{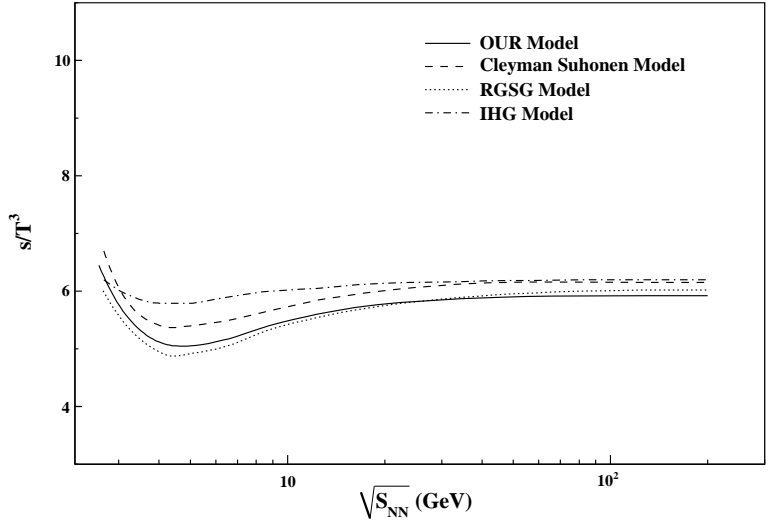}%use pdflatex for pdf
\caption[]{Variation of $s/T^{3}$ with $\sqrt{S_{NN}}$.Ideal HG model calculation is shown by dash dotted line,Cleymans-Suhonen and Rischke et al. model (RGSG model) calculation shown by dashed and dotted line respectively.Solid line shows the calculation by our model}
\end{figure}

Should normalized entropy density $s/T^3$ remain fixed over the entire chemical freeze-out surface in heavy-ion collision experiments ? This idea was initially used to extrapolate lattice gauge results from $\mu_B=0$ to finite values of $\mu_B$ by keeping $s/T^{3}$ fixed [47]. In Ref. [49] this quantity was also used to separate a baryon-dominant region from a meson-dominant one, in order to understand the rapid variations of certain particle ratios observed at lower SPS energies by the NA49 collaboration [29]. In Fig. 18, we have shown the energy dependence of normalized entropy density  $s/T^3$ which has energy dependence at lower energies in almost all the models. In Ideal HG model, however, energy independence was observed to some extent and its value equals to approximately 6.0. In case of Cleymans-Suhonen and RGSG model calculations, this ratio follows the same trend as in our model calculation. This ratio varies very wildly at lower energies in these excluded-volume models. Thus we cannot treat the criterion of fixed $s/T^{3}$ valid on the freeze-out hypersurface of the fireball as it is dependent on the energy of the collisions.   

\begin{figure}
\includegraphics[height=20em]{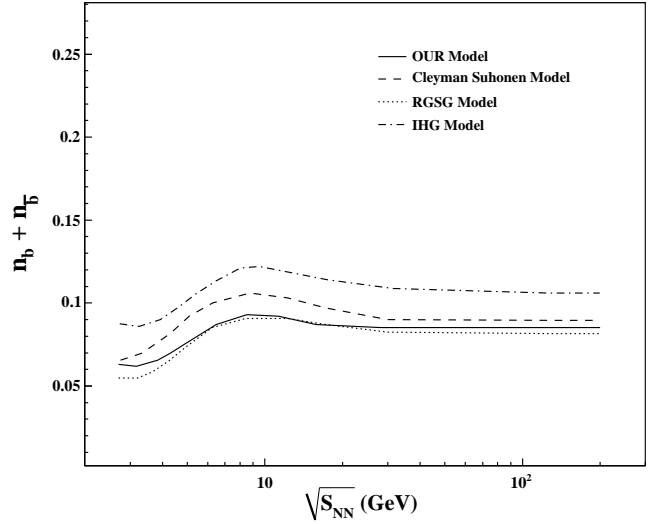}%use pdflatex for pdf
\caption[]{Variation of $n_{B}+n_{\bar{B}}$ with $\sqrt{S_{NN}}$.Ideal HG model calculation is shown by dash dotted line, Cleymans-Suhonen and Rischke et al. (RGSG model) model calculation shown by dashed and dotted line respectively.Solid line shows the calculation by our model}
\end{figure}

In heavy-ion collisions, the net  baryon density i.e. the difference between the density of baryons  $n_{B}$ and the density of anti-baryons $n_{\bar{B}}$ shows a very wild variation with the centre-of-mass energy as shown in Fig. 15. However, it was first noticed by Braun-Munzinger et al. [45] that the sum of baryon and anti-baryon densities remains constant at the chemical freeze-out. However, they have used the excluded-volume model of RGSG and they used different eigenvolumes for baryons and mesons, respectively. In Fig. 19, we have shown the variation of $n_{B}+n_{\bar{B}}$ with $\sqrt{S_{NN}}$. This quantity indeed involves a rapid variation with the energy in almost all the hadron gas models. Our calculations thus reveal that some of the above criteria are not strictly valid on the freeze-out surface as they do not show energy independence. However, as pointed out by certain authors, we can still treat them as freeze-out criteria provided we give adjustable eigenvolumes for both baryons and mesons, respectively. Our finding lends support to the crucial assumption of achieving chemical equilibrium by HG resulting in heavy-ion collisions from the lowest SIS upto RHIC energy and the EOS of the hadron-gas developed by us gives a proper description of the hot and dense fireball and its subsequent expansion. However, we still do not get any information regarding QGP formation from these criteria. The chemical equilibrium once attained by the hot and dense HG, removes any memory regarding QGP existing in the HG fireball. Furthermore, in a heavy-ion collision, a large amount of kinetic energy becomes available and part of it is always lost during the collision due to dissipative processes. In thermal description of the fireball, we ignore the effect of such processes and we assume that all available kinetic energy (or momentum) is globally thermalized at the freeze-out density. Experimental configuration of the collective flow in the hot, dense matter reveals the unsatisfactory nature of the above assumption.

\section{Transport Properties}
Transport coefficients are of particular interest to quantify the properties of strongly interacting relativistic fluid and its critical phenomena i.e., phase transition, critical point etc. [50-52]. The fluctuations cause the system to depart from equilibrium and a non-equilibrated system for a brief time is created. The response of the system to such fluctuations is essentially described by the transport coefficients e.g., shear viscosity, bulk viscosity, speed of sound etc. Recently the data for the collective flow obtained from RHIC and LHC experiments indicate that the system created in these experiments behaves as strongly interacting perfect fluid [53], whereas we expected that QGP created in these experiments should behave like a perfect gas. The perfect fluid created after the phase transition thus has a very low value of shear viscosity to entropy ratio so that the dissipative effects are negligible and the collective flow should be large as obtained by heavy ion collision experiments [54, 55]. There were several analytic calculations for $\eta$ and $\eta/s$ of simple hadronic systems [56-62] along with some sophisticated microscopic transport model calculations [63-65] in the literature. Furthermore, some calculations predict that the minimum of shear viscosity to entropy density is related with the QCD phase transition [66-70]. Similarly sound velocity is very important property of the matter created in heavy ion collision experiments because the hydrodynamic evolution of this matter strongly depends on it. A minimum in the sound-velocity has also been interpreted in terms of a phase transition [67, 71-77] and further, the presence of a shallow minimum corresponds to a cross-over transition [78]. Similarly Liao and Koch have shown that the shear viscosity to entropy ratio cannot give a good measure of fluidity when one compares the relativistic vis-a-vis non-relativistic fluid and they have defined a new fluidity variable for this purpose [79]. In view of the above, it is worthwhile to study in detail the transport properties of the HG in order to fully comprehend the nature of the matter created in the colliders as well as the involved phase transition phenomenon. In this section, our excluded volume model for HG has been used to calculate the transport properties like shear viscosity to entropy ratio, speed of sound and also the fluidity measure as proposed by Liao and Koch [79].  

Our calculation for the shear viscosity is completely based on the method of Gorenstein et al. [80] where RGSG model was used for HG. According to molecular kinetic theory, we can write the dependence of the shear viscosity as follows [81]:
\begin{equation}
\eta \propto n\;l\;\langle|{\bf p}|\rangle , 
\end{equation}
where $n$ is the particle density, $l$ is the mean free path, and hence the average thermal momentum of the baryons or antibaryons is:
\begin{equation}
\langle|{\bf p}|\rangle= \frac{\int_{0}^{\infty}p^{2}\;dp \;p \;{\bf A}}{\int_{0}^{\infty}p^{2}\;dp\;{\bf A}}, 
\end{equation}
and ${\bf A}$ is the Fermi-Dirac distribution function for baryons (anti-baryons). 
For the mixture of particle species with different masses and with the same hard-core radius $r$, the shear viscosity can be calculated by the following equation [80]:
\begin{equation}
\eta=\frac{5}{64 \sqrt{8} \;r^2}\sum_{i}\langle|{\bf{p}_{i}}|\rangle\times \frac{n_{i}}{n},
\end{equation}
where $n_{i}$ is the number density of the ith species of baryons (anti-baryons) and $n$ is the total baryon density.

In Fig.20, we have shown the variation of $\eta/s$ with respect to temperature as obtained in our model for HG having a baryonic hard-core size $r=0.5$ fm, and compared the results with those of Gorenstein et. al. [80]. We find that near the expected QCD phase transition temperature ($T_{c}=170-180$ MeV), $\eta/s$ shows a lower value in our HG model than the value in other model. In fact, $\eta/s$  in our model looks close to the lower bound ($1/4 \pi)$ suggested by AdS/QCD theories [82]. Recently measurements in Pb-Pb collisions at the Large Hadron Collider (LHC) support the value $\eta/s\approx 1/4 \pi$ when compared with the viscous fluid hydrodynamic flow [83].
\begin{figure}
\begin{center}
\includegraphics[height=15em]{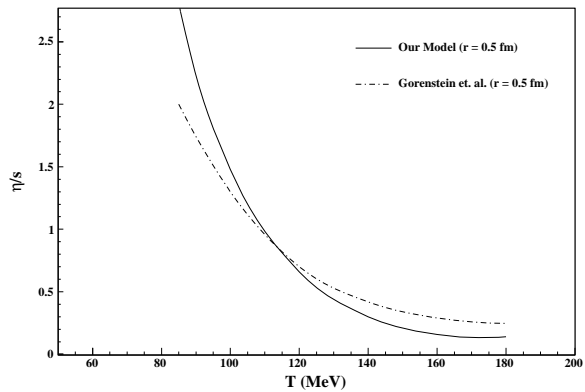}%use pdflatex for pdf
\caption[]{Variation of $\eta/s$ with temperature for $\mu_{B}=0$ in our model and a comparison with the results obtained by Gorenstein et. al. [80].}
\end{center}
\end{figure}
\begin{figure}
\begin{center}
\includegraphics[height=15em]{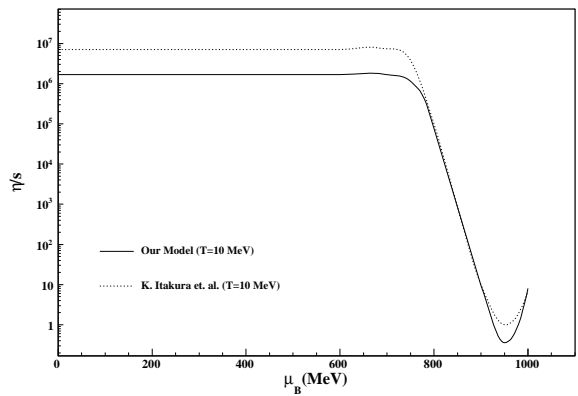}%use pdflatex for pdf
\caption[]{Variation of $\eta/s$ with  respect to baryon chemical potential ($\mu_{B}$) at very low temperature 10 MeV. Solid line represents our calculation and dotted curve is by K. Itakura et. al. [61].}
\end{center}
\end{figure}
\begin{figure}
\begin{center}
\includegraphics[height=20em]{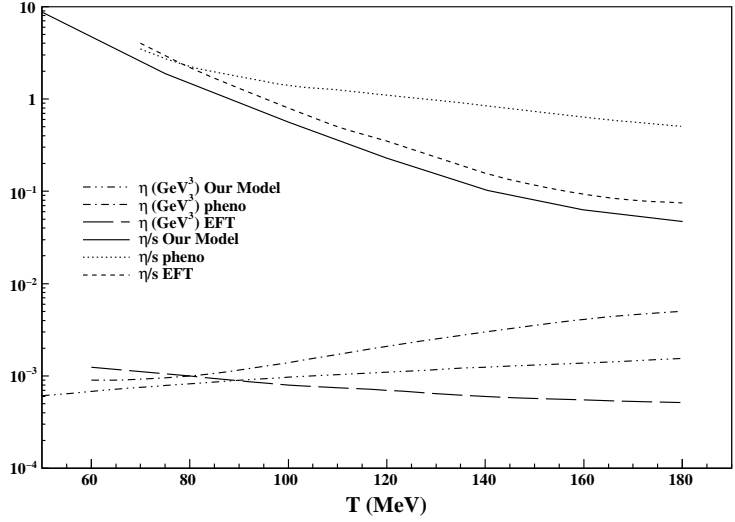}%use pdflatex for pdf
\caption[]{Variation of $\eta$ in unit of $(GeV)^3$ and $\eta/s$ with respect to temperature at $\mu_{B}=300 MeV$ in our model and a comparison with the results obtained by K. Itakura et. al. [61].}
\end{center}
\end{figure}
\begin{figure}
\begin{center}
\includegraphics[height=15em]{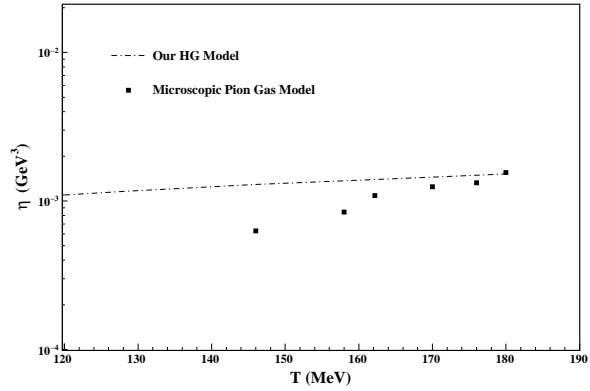}%use pdflatex for pdf
\caption[]{Variation of $\eta$ with respect to temperature at $\mu_{B}=300 MeV$ in our model and a comparison with the results obtained by A. Muronga [63].}
\end{center}
\end{figure}

In Fig.21, we have shown the variation of $\eta/s$ with respect to $\mu_{B}$ but at a very low temperature ($\approx 10$ MeV). Here we find that the $\eta/s$ is constant as $\mu_{B}$ increases upto $700$ MeV and then sharply decreases. This kind of valley-structure at low temperature and at $\mu_{B}$ around $950$ MeV was also obtained by J. W. Chen et al. and K. Itakura et. al. [59, 61]. They have related this structure to the liquid-gas phase transition of the nuclear matter. As we increase the temperature above $20$ MeV, then this valley-like structure disappears. They further suspect that the observation of a discontinuity in the bottom of $\eta/s$ valley may correspond to the location of the critical point. Our HG model yields a curve in complete agreement with these results. 

In Fig.22, we have shown the variation of $\eta$ and $\eta/s$ with respect to temperature at a fixed $\mu_{B}$ ($=300$ MeV), for HG having a baryonic hard-core size $r=0.8$ fm, and compared this result with the result obtained in Ref. [61]. Here we find that $\eta$ increases with temperature in our HG model as well as in the simple phenomenological calculation of Ref. [61], but in low temperature effective field theory (EFT) calculations, $\eta$ decreases with increase in temperature [59,61]. However, $\eta/s$ decreases with increasing temperature in all three calculations and $\eta/s$ in our model gives the lowest value at all the  temperatures in comparison to other models. 

In Fig.23, we have shown a comparison between $\eta$ calculated in our HG model with the results obtained in a microscopic pion gas model used in Ref [63]. Our model results show a fair agreement with the microscopic model results for the temperature higher than $160$ MeV while at lower temperatures the microscopic calculation predicts lower values of  $\eta$  in comparison to our results. The most probable reason may be that the calculations have been done only for pion gas in the microscopic model while at low temperatures the inclusion of baryons in the HG is very important in order to extract a correct value for the shear viscosity.

The speed of sound is another important quantity because it is related to the speed of small perturbations produced in the medium in its local rest frame. Here we have used the recent formulation of Cleymans and Worku to calculate the speed of sound at constant $s/n$ [72]. The speed of sound at zero chemical potential is easy to calculate where it is sufficient to keep the temperature constant [71, 77]. However, the speed of sound $(c_{s})$ at finite chemical potential can be obtained by using the following extended expression [72]:

\begin{equation}
c_{s}^{2}=\frac{\left(\frac{\partial p}{\partial T} \right)+ \left(\frac{\partial p} {\partial \mu_{B}} \right)\left(\frac{d\mu_{B}}{dT} \right)+\left(\frac{\partial p} {\partial \mu_{s}} \right)\left(\frac{d\mu_{s}}{dT} \right)}{\left(\frac{\partial \epsilon}{\partial T} \right)+ \left(\frac{\partial \epsilon} {\partial \mu_{B}} \right)\left(\frac{d\mu_{B}}{dT} \right)+\left(\frac{\partial \epsilon} {\partial \mu_{s}} \right)\left(\frac{d\mu_{s}}{dT} \right)},
\end{equation}
where the derivative $d\mu_{B}/dT$ and $d\mu_{s}/dT$ can be evaluated by using two conditions, firstly of keeping $s/n$ constant and then imposing overall strangeness neutrality. Thus one gets [72]:
\begin{equation}
\frac{d\mu_{B}}{dT}=\frac{\left[n\left(\frac{\partial s}{\partial \mu_{s}}\right)-s\left(\frac{\partial n}{\partial \mu_{s}}\right) \right] \left[\frac{\partial L}{\partial T}-\frac{\partial R}{\partial T}\right] -\left[n\left(\frac{\partial s}{\partial T}\right)-s\left(\frac{\partial n}{\partial T}\right)\right]\left[\frac{\partial L}{\partial \mu_{s}}-\frac{\partial R}{\partial \mu_{s}}\right]} {\left[n\left(\frac{\partial s}{\partial \mu_{B}}\right)-s\left(\frac{\partial n}{\partial \mu_{B}}\right)\right]\left[\frac{\partial L}{\partial \mu_{s}}-\frac{\partial R}{\partial \mu_{s}}\right] -\left[n\left(\frac{\partial s}{\partial \mu_{s}}\right)-s\left(\frac{\partial n}{\partial \mu_{s}}\right)\right]\left[\frac{\partial L}{\partial \mu_{B}}-\frac{\partial R}{\partial \mu_{B}}\right]},
\end{equation}
and
\begin{equation}
\frac{d\mu_{s}}{dT}=\frac{\left[n\left(\frac{\partial s}{\partial T}\right)-s\left(\frac{\partial n}{\partial T}\right) \right] \left[\frac{\partial L}{\partial \mu_{B}}-\frac{\partial R}{\partial \mu_{B}}\right] -\left[n\left(\frac{\partial s}{\partial \mu_{B}}\right)-s\left(\frac{\partial n}{\partial \mu_{B}}\right)\right]\left[\frac{\partial L}{\partial T}-\frac{\partial R}{\partial T}\right]} {\left[n\left(\frac{\partial s}{\partial \mu_{B}}\right)-s\left(\frac{\partial n}{\partial \mu_{B}}\right)\right]\left[\frac{\partial L}{\partial \mu_{s}}-\frac{\partial R}{\partial \mu_{s}}\right] -\left[n\left(\frac{\partial s}{\partial \mu_{s}}\right)-s\left(\frac{\partial n}{\partial \mu_{s}}\right)\right]\left[\frac{\partial L}{\partial \mu_{B}}-\frac{\partial R}{\partial \mu_{B}}\right]},
\end{equation}
where $L=n_{s}^{B}+n_{s}^{M}$, is the sum of the strangeness density for baryons and mesons. Similarly $R=n_{s}^{\bar{B}}+n_{s}^{\bar{M}}$, the sum of anti-strangeness density for baryons and mesons.

In Fig.24, we have shown the variation of $c_{s}^2$ with respect to $\mu_{B}$ at two different temperatures. We find that at $T=120$ MeV, there is a clear minimum at $\mu_{B}\approx 500$ in the curve of the speed of sound while in the case of $T=170$ MeV, we do not observe any such minimum and $c_{s}^2$ continues increasing with increase in $\mu_{B}$. The minimum at $\mu_{B}\approx 500$ for $T=120$ MeV indicates the position where a first order phase transition from HG to QGP materializes.

\begin{figure}
\begin{center}
\includegraphics[height=15em]{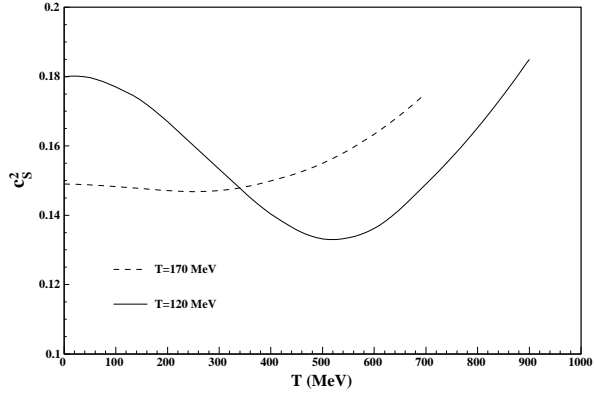}%use pdflatex for pdf
\caption[]{Variation of square of speed of sound with respect to $\mu_{B}$ at different temperatures.}
\end{center}
\end{figure}
In a recent paper [79], Liao and Koch have suggested that $\eta/s$ can serve as a good measure for the fluidity of a relativistic fluid only because the ability of $\eta/s$ to serve such a role is actually inherited from $\eta/\omega$ where the enthalpy of HG is $\omega$. We find that $\omega$ becomes approximately equal to $Ts$ only in the case of relativistic or ultrarelativistic matter. It is not always necessary that one can prefer $\eta/s$ in place of $\eta/\omega$ for a measure of the fluidity of the system. Thus if we want to compare various systems i.e. relativistic and non-relativistic, and extract some useful insights about the nature of the system then one has to define a fluidity measure exclusively in terms of the properties of the system itself. Liao and Koch defined a quantity $F$ to measure the fluidity of the relativistic and/or non-relativistic system as follows :
\begin{equation}
F=\frac{L_{\eta}}{L_{n}},
\end{equation}
where we can use
\begin{equation}
L_{\eta}=\frac{\eta}{\omega c_{s}},
\end{equation}
and $L_{n}$ can be calculated by the following relation [79]:
\begin{equation}
L_{n}= \frac{1}{n^{1/3}}=\left(\frac{4}{s}\right)^{1/3}.
\end{equation}
Actually $L_{\eta}$ provides a measure for the minimal wavelength of a sound wave which propagates in a viscous fluid and $L_{n}$ is basically related with the inter-particle distance to provide an internal length-scale for the medium. In Fig. 25, we have shown the variation of $F=\frac{L_{\eta}}{L_{n}}$ with respect to temperature as obtained in our excluded-volume model using different hard-core sizes for the baryons and compared the results with the curve obtained by Liao and Koch [79] in which they crudely assumed $\eta/T_{c}^{3} \approx T/T_{c}$. We thus find that the features of our curves give similar behaviour as the results obtained by Liao and Koch using altogether a completely different formalism. 

The study of the transport properties of non-equilibrium systems which are not far from an equilibrium state has yielded valuable results in the recent past. Large values of the elliptic flow observed at RHIC indicates that the matter in the fireball behaves as a nearly perfect liquid with a small value of the $\eta/s$ ratio. After evaluating $\eta/s$ in strongly coupled theories using AdS/CFT duality conjecture, a lower bound was reported as $\eta/s=\frac{1}{4\pi}$. We surprisingly notice that the fireball with hot, dense HG as described in our excluded-volume model gives transport coefficient which agree with those given in different approaches. Temperature and baryon chemical potential dependence of the $\eta/s$ are analyzed and compared with the results obtained in other models. We also focus our attention to $c_s$ and the fluidity variable. Our results show the similar trends and features as have been reported by previous authors.      

\begin{figure}
\begin{center}
\includegraphics[height=15em]{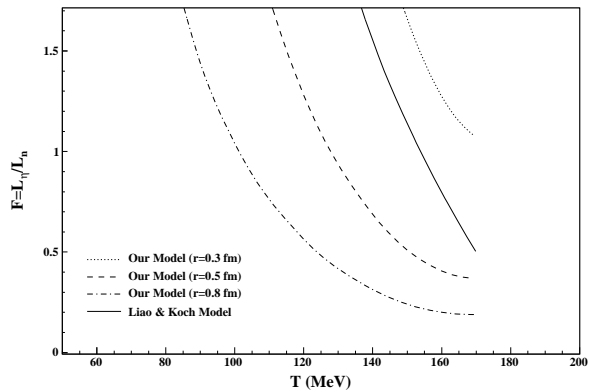}%use pdflatex for pdf
\caption[]{Variation of fluidity (F) with temperature for $\mu_{B}=0$ in our model and a comparison with the results obtained from Liao and Koch [79]. }
\end{center}
\end{figure}

\section{Summary and Conclusions}
We have formulated a new thermodynamically consistent EOS for a hot and dense HG by incorporating a hard-core finite size of the baryons and antibaryons only. We have treated mesons as pointlike particles. Alternatively they possess a size but they can penetrate and overlap on each other. Our prescription is valid even at extreme values of $T$ and $\mu_B$. Moreover, our model does not suffer from either of the two main inconsistencies i.e., violation of causality as well as thermodynamic inconsistency. Our model involves a mathematical form which resembles with the thermodynamically inconsistent Cleymans-Suhonen model but contains some extra correction terms which arise due to the condition of thermodynamic consistency. We have calculated the prediction of our model for various thermodynamic quantities like pressure, energy density, number density, entropy density etc. and compared the predictions with those of other excluded-volume models. Similarly we have also compared our results with those obtained from a microscopic simulation approach of Sasaki. We find that our results mostly show very close agreements with those of Sasaki, although the two approaches are completely different in nature. In addition, Sasaki's approach has a fundamental inconsistency and anti-baryons and strange particles are not included in the model. The EOS thus formulated usually suffers from a crucial assumption regarding how many particles and resonances one should incorporate into it. We have taken all the known particles and resonances upto the mass of $2 GeV/c^2$.

Our results for the particle ratios and their energy dependences fit the experimental data very well. We have deduced certain freeze-out criteria and attempted to test whether these criteria involve energy-independence as well as independence of the structures of the nuclei involved in the collision. We find that two criteria, i.e. $E/N=1.0 GeV$ per produced particle and $s/n=7.0$ demonstrate their validity. Moreover, the calculations of transport properties in our model match well with the results obtained in other widely different approaches.

In conclusion, the utility of our present model has been demonstrated in explaining various properties of hot, dense hadron gas and thus our model provides a proper and realistic EOS for a hot, dense HG and it can suitably describe HG at extreme values of temperatures and/or densities. The calculations regarding $p_T$ as well as the rapidity spectra of different particles at RHIC and LHC are still in progress and it will appear in a future paper.

\section{Acknowledgments}
SKT and PKS are grateful to Council of Scientific and Industrial Research (CSIR), New Delhi and University Grants Commission (UGC) for providing a research fellowship. CPS acknowledges the financial support through a project sanctioned by Department of Science and Technology, Government of India, New Delhi.

\end{document}